\documentclass[fleqn,10pt]{wlscirep}
\usepackage{graphicx}
\usepackage{epsfig}
\usepackage{epstopdf}
\usepackage{amsmath}

\title{Hybridization oscillation in the one-dimensional Kondo-Heisenberg model with Kondo holes}

\author[1,2,$\dagger$]{Neng Xie}
\author[1,2,$\dagger$]{Danqing Hu}
\author[1,2,3,*]{Yi-feng Yang}
\affil[1]{Beijing National Laboratory for Condensed Matter Physics and Institute of Physics, Chinese Academy of Sciences, Beijing 100190, China}
\affil[2]{School of Physical Sciences, University of Chinese Academy of Sciences, Beijing 100190, China}
\affil[3]{Collaborative Innovation Center of Quantum Matter, Beijing 100190, China}

\affil[$\dagger$]{These authors contributed equally to this work.}
\affil[*]{yifeng@iphy.ac.cn}

\begin{abstract}
We use the density matrix renormalization group method to study the properties of the one-dimensional Kondo-Heisenberg model doped with Kondo holes. We find that the perturbation of the Kondo holes to the local hybridization exhibits spatial oscillation pattern and its amplitude decays exponentially with distance away from the Kondo hole sites. The hybridization oscillation is correlated with both the charge density oscillation of the conduction electrons and the oscillation in the correlation function of the Heisenberg spins. In particular, we find that the oscillation wavelength for intermediate Kondo couplings is given by the Fermi wavevector of the large Fermi surface even before it is formed. This suggests that heavy electrons responsible for the oscillation are already present in this regime and start to accumulate around the to-be-formed large Fermi surface in the Brillouin zone.
\end{abstract}
\begin{document}

\flushbottom
\maketitle

\thispagestyle{empty}

Heavy fermion materials exhibit many exotic quantum phenomena such as unconventional superconductivity \cite{Steglich1979,Pfleiderer2009,White2015} and unconventional quantum criticality \cite{Coleman2001,Si2001,Stockert2011}. The existence of these exotic quantum states is closely related to the host state, namely the heavy electron Kondo liquid, which emerges below a characteristic temperature, $T^*$, due to the collective hybridization of a lattice of localized $f$-spins with the conduction electron sea \cite{Yang2008,Yang2012}. While the so-called Kondo problem of a single localized moment antiferromagnetically coupled to the conduction electrons has been well understood, the Kondo lattice problem remains controversial and poses a long standing challenge for the condensed matter community \cite{Hewson1993,Coleman2007,Yang2016}. The difficulty lies in the lack of a good understanding of the collective nature of the underlying spin entanglement or hybridization between the localized spins and the conduction electrons \cite{Lonzarich2016}. Recently, it was realized that by introducing local Kondo holes, or defects/nonmagetic impurities in the lattice of the local moments \cite{Figgins2011,Balatsky2012,Baruselli2014}, it is possible to stimulate a collective spatial modulation in the hybridization strength which can be probed by using the state-of-the-art spectroscopic imaging scanning tunneling microscopy (STM) \cite{Hamidian2011,Yazdani2016}. Theoretical calculations based on the mean-field approximation predicted a spatial oscillation of the hybridization with a characteristic wavelength determined by the Fermi wavevector of the so-called small Fermi surface of unhybridized conduction electrons \cite{Figgins2011}, which seems to be confirmed by later STM experiment on Th-doped URu$_2$Si$_2$ at very low temperature in the hidden order phase \cite{Hamidian2011}. 

Bearing in mind the limitation of the mean-field approximation, we examine the above results by applying the density matrix renormalization group (DMRG) method \cite{White1992,White1993,Schollwock2005} to the one-dimensional (1D) Kondo-Heisenberg model doped with Kondo holes. This allows us to solve the model exactly and take into fully account magnetic quantum fluctuations that are beyond the mean-field approximation \cite{Yang2008,Yang2012}. Our results confirm the predicted hybridization oscillation induced by the Kondo holes. However, contrary to the mean-field prediction, the characteristic wavelength of the oscillation changes dramatically from weak coupling to strong coupling regimes. For both intermediate and strong couplings, we find that the local hybridization, the conduction electron charge density, and the correlation function of the Heisenberg spins are all entangled and exhibit similar oscillation pattern. In particular, away from half filling, we find that the oscillation wavelength is determined by the Fermi wavevector of the large Fermi surface even before it is formed, distinctly different from previous mean-field predictions for the charge density oscillation. This indicates that preformed heavy electrons are already present and start to accumulate around the to-be-formed large Fermi surface at intermediate couplings. Our results suggest that these emergent heavy electrons can be detected by the Kondo hole induced hybridization oscillation using the scanning tunneling spectroscopy.

\begin{figure}[ht]
\centering
\resizebox{10cm}{!}{\includegraphics{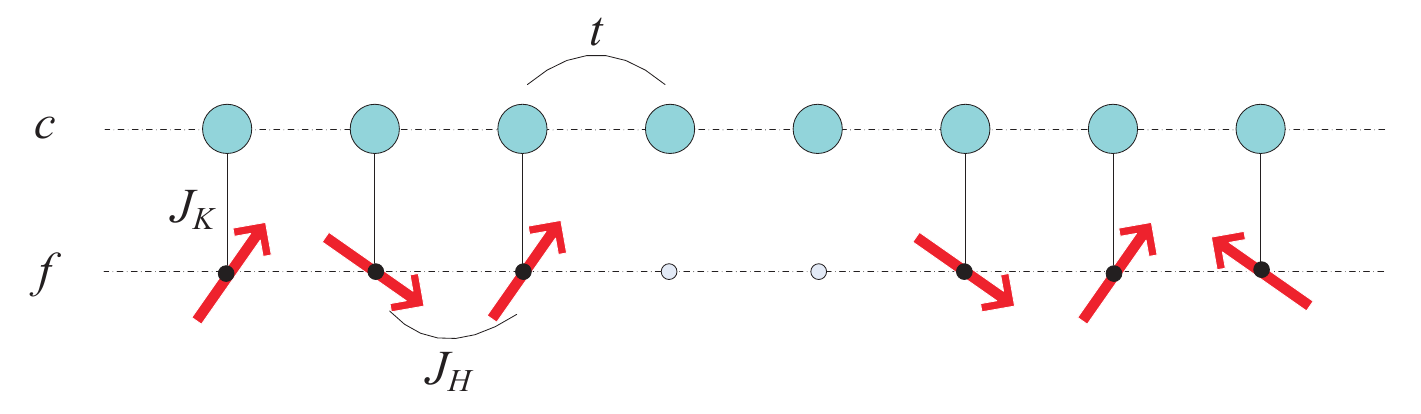}}
\caption{An illustration of the one-dimensional Kondo-Heisenberg model with Kondo holes in the middle of the spin chain.
\label{fig1}}
\end{figure}

\section*{Results}
We start with the following Hamiltonian for the 1D Kondo-Heisenberg model with Kondo holes, 
\begin{equation}
H= -t\sum_{i,\sigma}\left(c_{i,\sigma}^{\dagger}c_{i+1,\sigma}+{\rm H.c.} \right)+J_K\sum_{i}{}^\prime\vec{S}_{i}\cdot\vec{s}_{i}+J_H\sum_{i}{}^\prime\vec{S}_{i}\cdot\vec{S}_{i+1},
\label{hamiltonian}
\end{equation}
where the sum ($\sum^\prime$) is over all spin sites other than the hole sites located in the middle of the Heisenberg chain as illustrated in Fig. \ref{fig1}. $c_{i,\sigma}^{\dagger }$($c_{i,\sigma}$) is the creation (annihilation) operator of the conduction electron with spin $\sigma$ at the $i$-th site ($i=1,\cdots,L$) and $\vec{s}_{i}=\sum_{\sigma,\sigma^\prime}c_{i,\sigma}^{\dagger }(\vec{\sigma}/2)_{\sigma,\sigma^\prime}c_{i,\sigma^\prime}$, where $\vec{\sigma}$ are the Pauli matrices, defines the spin density operator. $\vec{S}_{i}$ denotes the local Heisenberg spin at the $i$-th site. $t$ is the hopping parameter of the conduction electrons between neighboring sites, and $J_K/t>0$ is the local Kondo coupling. We further introduce a finite antiferromagnetic exchange coupling $J_H/t=0.5$ between nearest-neighbor Heisenberg spins to avoid possible ferromagnetic ground state away from half filling \cite{Tsunetsugu1997,Gulacsi2004,Moukouri1996,Xie2015}. For numerical simplicity, we remove two local spins ($i_h=L/2, L/2+1$) to keep the inversion symmetry and a nonmagnetic ground state. The results are similar but less pronounced if only one local spin is removed. The model is calculated with a modified {\scriptsize{DMRG++}} code \cite{Alvarez2009} using $800$ block states for $L=100$ sites and with open boundary condition. The presented results have been verified to converge with different numbers of lattice sites and block states and also examined using the exact diagonalization method on a lattice of up to 14 sites with both periodic and open boundary conditions. Similar conclusions are also obtained in a ladder system with $L=50$. We consider the ground state $S^z_{tot}=\sum_i s_i^z+ \sum^\prime_i S_i^z=0$ for each fixed average occupation number of the conduction electrons, $n^c=L^{-1}\sum_{i\sigma}c_{i\sigma}^\dagger c_{i\sigma}$. To avoid the boundary effect \cite {Shibata1996,Xavier2002}, we subtract the results of the corresponding clean Kondo lattice without the Kondo holes and study the changes in the local hybridization, $\delta\mathcal{V}_i=\langle \vec{S}_{i}\cdot\vec{s}_{i}\rangle-\langle \vec{S}_{i}\cdot\vec{s}_{i}\rangle_0$, the charge density of the conduction electrons, $\delta n^c_i=\sum_{\sigma}\left(\langle c_{i,\sigma}^{\dagger}c_{i,\sigma}\rangle- \langle c_{i,\sigma}^{\dagger}c_{i,\sigma}\rangle_0\right)$, and the correlation function of local spins, $\delta\chi_i=\langle \vec{S}_{i}\cdot\vec{S}_{i+1}\rangle-\langle \vec{S}_{i}\cdot\vec{S}_{i+1}\rangle_0$. The subscript "0" indicates the corresponding results for the clean Kondo lattice as a background for comparison.

\begin{figure}[ht]
\centering
\resizebox{10cm}{!}{\includegraphics{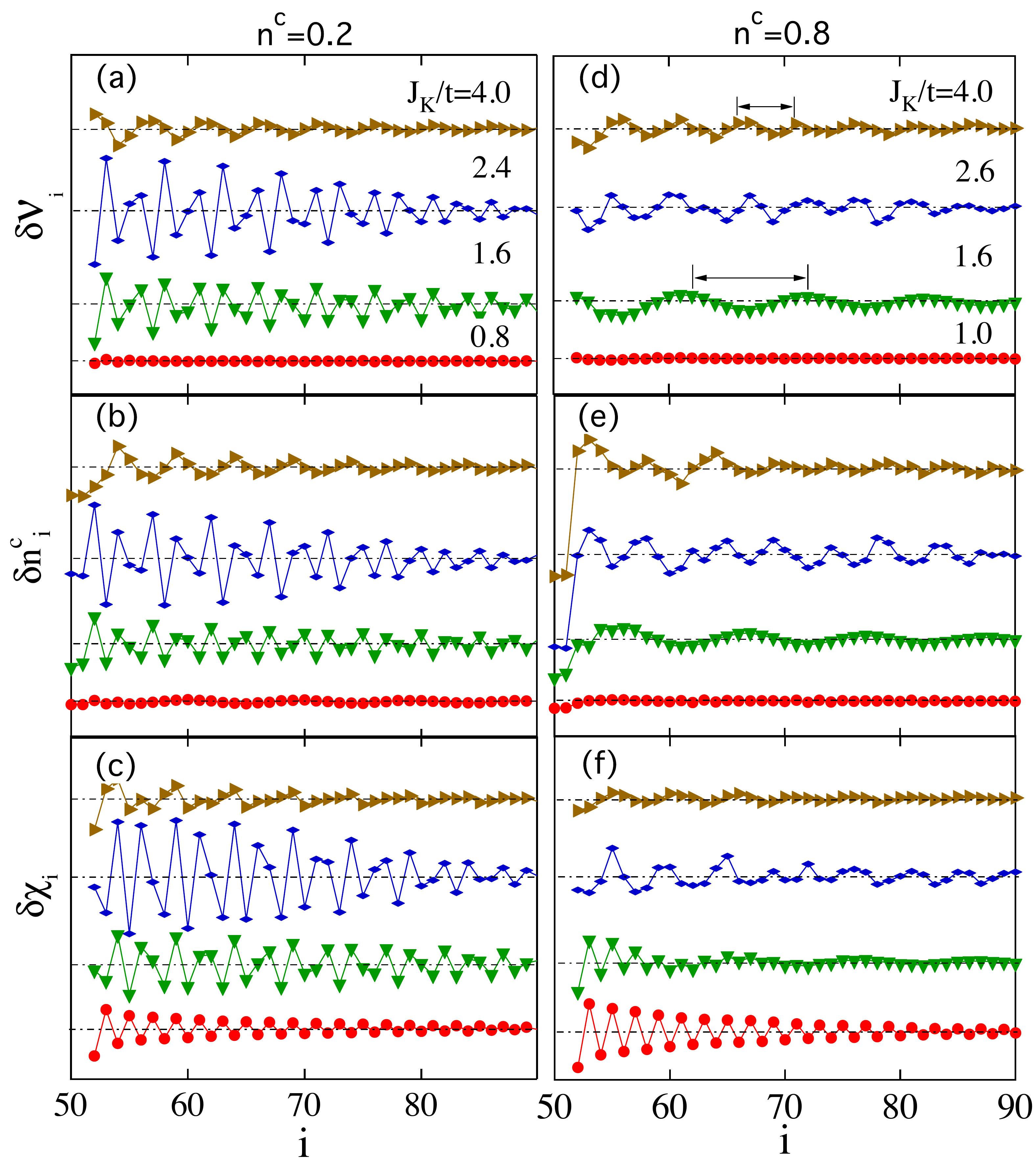}}
\caption{The variation of the spatial oscillation patterns induced by the Kondo holes for $\delta \mathcal{V}_i$, $\delta n^c_i$ and $\delta \chi_i$ with the Kondo coupling $J_K/t$. Other parameters are $J_H /t = 0.5$ and $L=100$ for both $n^c=0.2$ and $0.8$.
\label{fig2}}
\end{figure}

We first discuss the results for $n^c<1$, namely away from the half filling. Fig. \ref{fig2} presents some typical results with $n^c=0.2$ and 0.8 for different Kondo couplings, $J_K/t$. As expected, the resulting $\delta n^c_{i}<0$ at the Kondo hole sites is distinctly different from that induced by an attractive nonmagnetic impurity. We observe spatial oscillations in almost all the cases and the amplitude of the oscillation decays gradually with distance away from the hole sites. A straightforward comparison of the results at different $J_K/t$ indicates that the wavelength of the oscillation changes dramatically from the weak coupling regime to the strong coupling regime. Moreover, there seems to be a close correlation of the oscillation wavelength in all three quantities in the strong coupling regime. In the weak coupling regime, on the other hand, the oscillation is hardly seen in $\delta\mathcal{V}_i$, whereas the charge density $\delta n^c_i$ and the spin correlation, $\delta\chi_i$, also exhibit different oscillation patterns, suggesting that the conduction electrons are not strongly coupled with the local spins.

\begin{figure}[ht]
\centering
\resizebox{10cm}{!}{\includegraphics{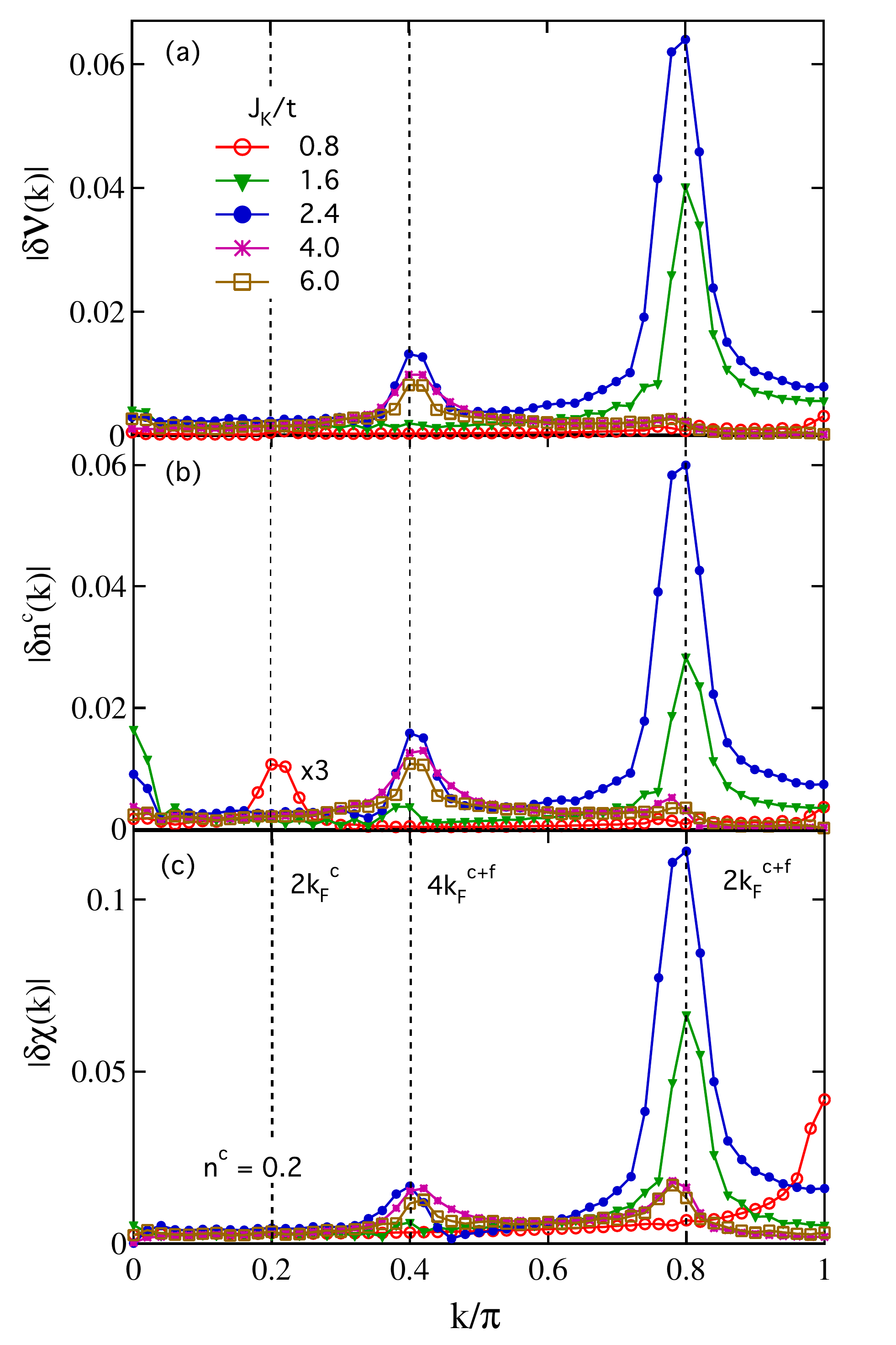}}
\caption{Absolute values of the Fourier transforms of $\delta \mathcal{V}_i$, $\delta n^c_i$ and $\delta \chi_i$ for $n^c=0.2$, $J_H/t=0.5$ and different Kondo couplings, $J_K/t$.
\label{fig3}}
\end{figure}

To see these results more clearly, we present in Fig.~\ref{fig3} the Fourier transforms of $\delta\mathcal{V}_i$, $\delta n^c_i$, and $\delta\chi_i$ in the momentum space for $n^c=0.2$, where the features are more pronounced. We see that all spectra exhibit peak structures. For $J_K/t=0.8$ in the weak coupling regime, $\delta\mathcal{V}(k)$ is negligible small with only a tiny peak at $k=\pi$. In contrast, $\delta\chi(k)$ shows a sharp peak at $k=\pi$ due to the strong inter-site antiferromagnetic correlations between neighboring spins, and $\delta n^c(k)$ has two peaks at $k=\pi n^c=2k_F^c$ and $k=\pi$ where the former corresponds to the small Fermi surface of the conduction electrons and the latter comes from the influence of the spin chain as is the case for the small peak at $k=\pi$ in $\delta\mathcal{V}_i$. The weak oscillations in both $\delta\mathcal{V}_i$ and $\delta\chi_i$ indicate that the conduction electrons and the local spins are effectively decoupled (or only very weakly coupled). For $J_K/t=1.6$ in the intermediate coupling regime, a sharp peak appears at $k=\pi(1-n^c)=2k_F^{c+f}\,({\rm mod}\,2\pi)$ in all three quantities. Here $k_F^{c+f}$ is the Fermi wavevector of the large Fermi surface, indicating that the conduction electrons are now coupled to the local spins and the large Fermi surface starts to take effect. For $J_K/t=2.4$, a new peak emerges at $k=2\pi n^c$, which corresponds to either $4k_F^{c}$ or $4k_F^{c+f} ({\rm mod}\,2\pi)$. Further increasing $J_K/t$ seems to suppress the peak at $2k_F^{c+f}$ but keep the peak at $4k_F^{c+f}$ almost unchanged. We note that the $4k_F^{c+f}$ or $4k_F^c$ oscillation may be understood as a special one-dimensional feature that originates from the spinless hole Fermi surface due to complete spin-charge separation following the formation of the Kondo singlets between the conduction electrons and the local spins \cite{Shibata1996}. Although we cannot distinguish $4k_F^{c}$ from $4k_F^{c+f}\,({\rm mod}\,2\pi)$ by number, we attempt to ascribe this peak to $4k_F^{c+f}\,({\rm mod}\,2\pi)$ associated with the large Fermi surface because of its coexistence with the $2k_F^{c+f}$ peak, whereas the $2k_F^c$ peak is absent in these regimes.

We note that the critical Kondo coupling is about $J_K/t\approx 3.0$ for $n^c=0.2$ and $J_H/t=0.5$ in the clean Kondo lattice \cite{Xie2015}, as determined from the maximum change in the distribution function of conduction electrons in the momentum space. It is quite unexpected that the hybridization oscillation is determined by $k_F^{c+f}$ at $J_K/t=1.6$ and 2.4 before the large Fermi surface is even formed. This implies that heavy electrons are already present in this intermediate coupling regime and start to accumulate around $k_F^{c+f}$ in the momentum space. Our results for $n^c<1$ suggest three regimes of the Kondo lattice physics depending on the magnitude of $J_K/t$: (1) the weak coupling regime where the conduction electrons and the local spins are effectively decoupled (or only very weakly coupled); (2) the strong coupling regime where the two components are coupled to give rise to a well-defined large Fermi surface and complete spin-charge separation in 1D (corresponding to the $4k_F^{c+f}$ peak); (3) the intermediate regime where the large Fermi surface is not yet formed, but preformed heavy electrons are already present and all three quantities, $\delta\mathcal{V}_i$, $\delta n^c_i$ and $\delta\chi_i$, exhibit similar oscillation pattern (corresponding to the $2k_F^{c+f}$ peak). These preformed heavy electrons are primarily responsible for the charge density and hybridization oscillations induced by the Kondo holes, whereas local spin singlets are not fully established so that the hybridization must be collective in nature and the global spin singlet state must involve highly nonlocal entanglement between neighboring spins and conduction electrons. 

We would like to point out that previous mean-field calculations predicted different periodicity for the three quantities, with $2k_F^c$ for $\delta n^c_i$ and $\delta\mathcal{V}_i$ and $2k_F^{c+f}$ for $\delta\chi_i$ in all parameter regimes \cite{Figgins2011}. This is in contradiction with our results where all three quantities have the same periodicity in the intermediate and strong coupling regimes. At the moment it is not clear what causes this discrepancy. It could be due to the one dimensional nature (with strong quantum fluctuations) of our DMRG calculations. However, our calculations on a ladder ($L=50$) yield similar oscillation patterns. Also, it is natural to imagine that all three quantities should be strongly entangled and have the same periodicity at least in the strong coupling regime. Moreover, the same $2k_F^{c+f}$ oscillation for the spin correlation $\delta\chi_i$ obtained in our approach and the mean-field approach seems to support the idea that our results may still be valid in higher dimensions. If this is the case, one may attempt to think that the mean-field calculations yield wrong predictions for treating incorrectly (as simple hybridization bands) the correlated states of collectively entangled conduction electrons and local spins. A thorough investigation of this discrepancy may lead to a deeper understanding of the various approaches to the Kondo lattice problem. Careful analysis and numerical calculations in higher dimensions will be crucial in order to clarify the role of quantum fluctuations in determining the oscillation pattern. As a consequence, the observation of the hybridization oscillation at the small Fermi wavevector in the hidden order phase of Th-doped URu$_2$Si$_2$ may probably need to be reinterpreted or reexamined in the normal state \cite{Hamidian2011}.

\begin{figure}[ht]
\centering
\resizebox{10cm}{!}{\includegraphics{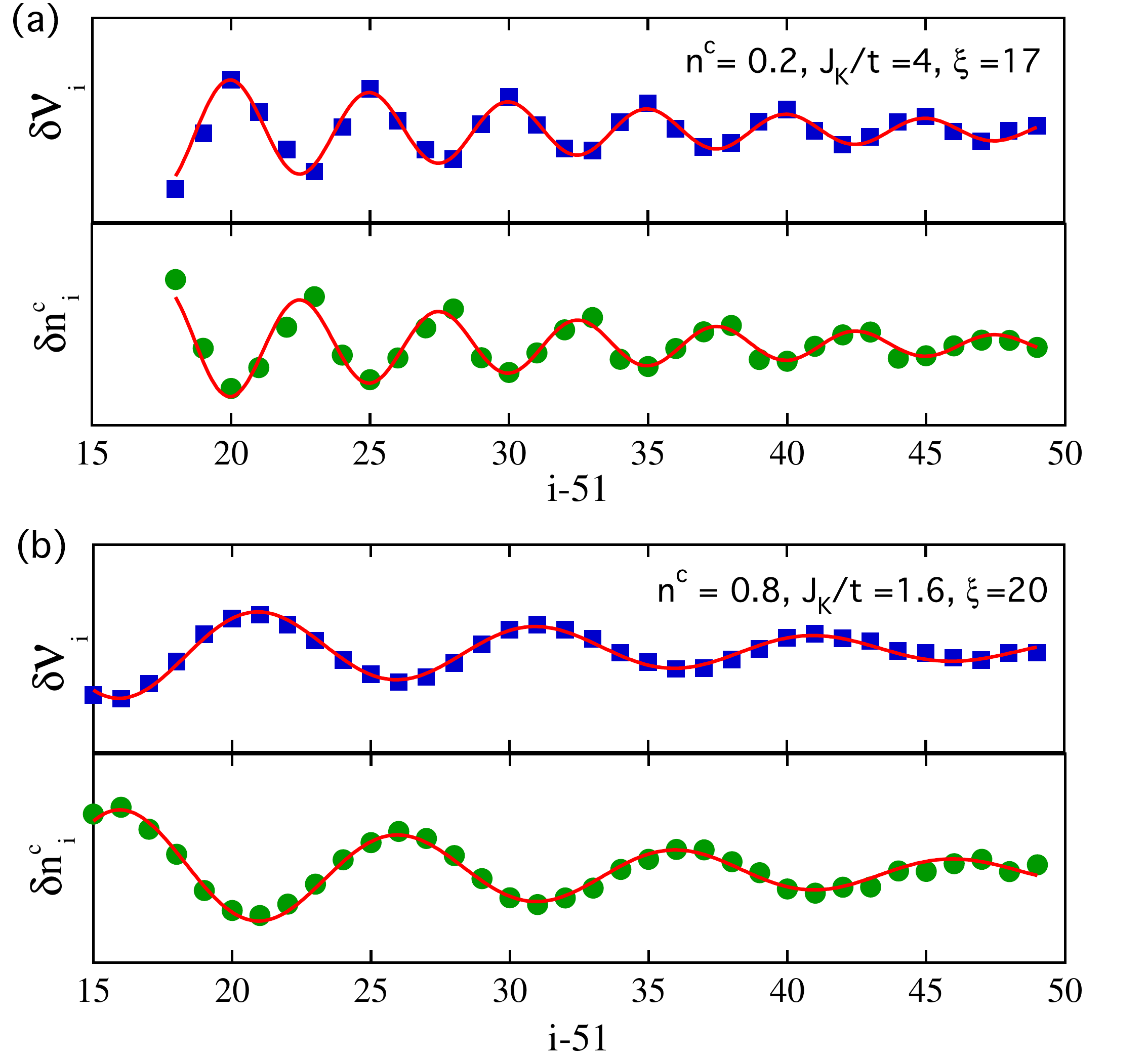}}
\caption{Fit to $\delta\mathcal{V}_i$ and $\delta n^c_i$ for (a) $n^c=0.2$ and $J_K/t=4$; (b) $n^c=0.8$ and $J_K/t=1.6$. The solid lines are the fitting curves.
\label{fig4}}
\end{figure}

To obtain a quantitative understanding of the spatial decay of the oscillation, we consider two examples where the spectra are governed by one dominant sharp peak in the momentum space. The correlation functions are then fitted using the following form in real space,
\begin{eqnarray}
\delta\mathcal{V}_i &=& A_\mathcal{V}\cos(2\pi x_i/\lambda+\theta_\mathcal{V}){\rm e}^{-x_i/\xi_\mathcal{V}},\nonumber\\
\delta n^c_i &=& A_n\cos(2\pi x_i/\lambda+\theta_n){\rm e}^{-x_i/\xi_n},
\end{eqnarray}
where the oscillation wavelength $\lambda$ is the same for the two quantities as discussed above and $x_i=i-L/2-1$ is the distance of the $i$-th lattice site from the Kondo holes. $\theta_{\mathcal{V}/n}$ is the phase shift and $\xi_{\mathcal{V}/n}$ denotes the characteristic decay length of the oscillation. Fig. \ref{fig4} shows two examples at $n^c=0.2$ and 0.8. We find an excellent agreement for both $\delta\mathcal{V}_i$ and $\delta n^c_i$. Both quantities exhibit similar decay lengths, $\xi_\mathcal{V}=\xi_{n}=\xi$. On the other hand, the oscillation in $\delta\chi_i$ in the strong coupling regime has a smaller decay length, despite that it exhibits the same periodicity. This suggests that the oscillation in $\delta\mathcal{V}_i$ is most affected by the charge density oscillation for $n^c<1$. We note that the above data do not follow a power law decay that is typically expected in 1D systems. This may originate from the finite Kondo energy and indicate that the hybridization perturbation is different from the usual Friedel oscillation.

\begin{figure}[ht]
\centering
\resizebox{10cm}{!}{\includegraphics{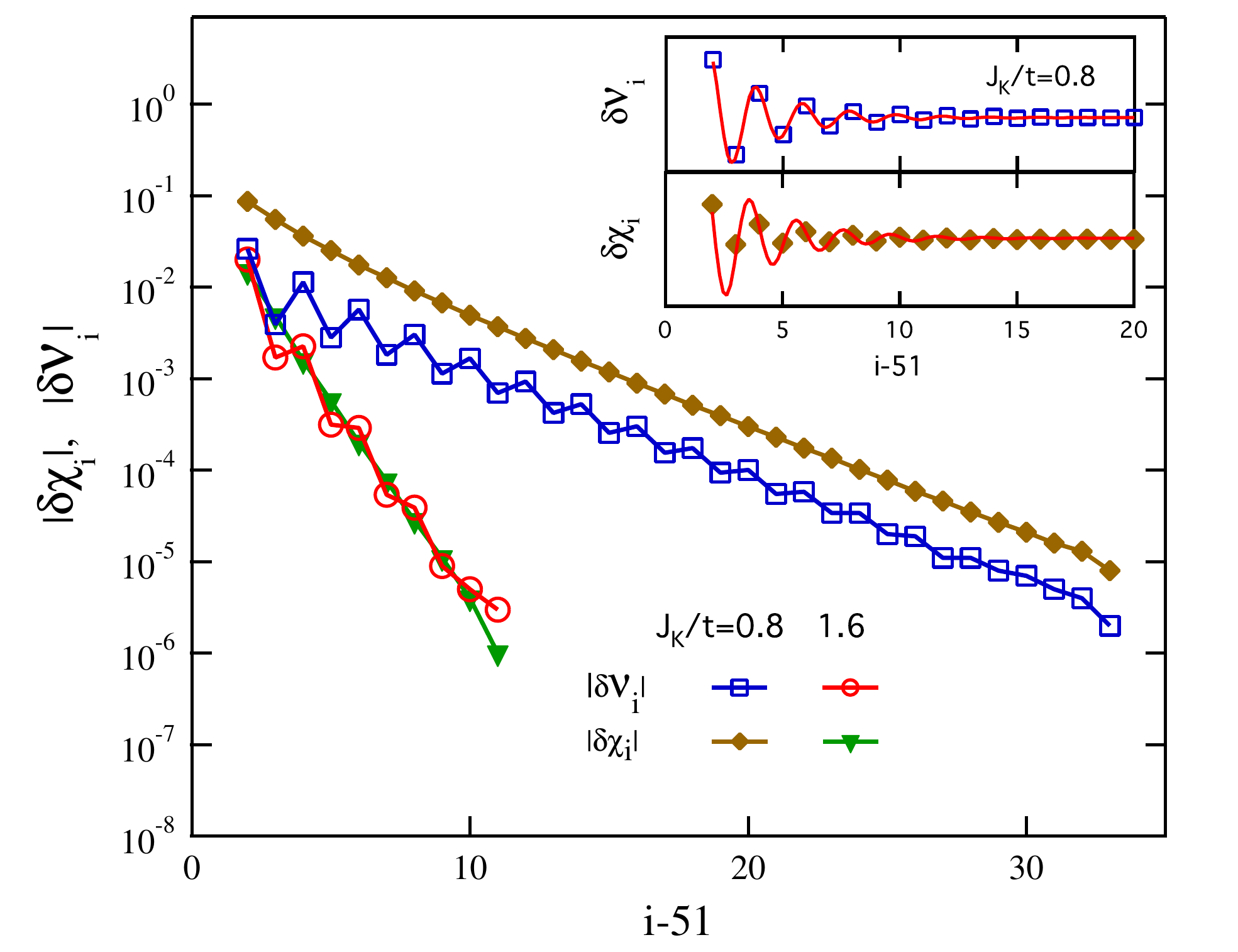}}
\caption{A semilog plot for the absolute values of $\delta\mathcal{V}_i$ and $\delta \chi_i$ as a function of the distance from the Kondo holes for $n^c=1$ and $J_K/t=0.8$, 1.6. The inset shows the original data, $\delta\mathcal{V}_i$ and $\delta \chi_i$, for $J_K/t=0.8$ on the linear scale.
\label{fig5}}
\end{figure}

Next we study briefly the half-filling case at $n^c=1$, where the density oscillation is suppressed due to the particle-hole symmetry and the finite charge gap. We therefore only need to consider the local hybridization $\delta\mathcal{V}_i$ and the spin correlation function $\delta\chi_i$. Fig. \ref{fig5} compares the two quantities on a logarithmic-linear scale as a function of the distance from the Kondo holes. We see that $|\delta\mathcal{V}_i|$ and $|\delta\chi_i|$ exhibit similar slope ($\propto \xi_{\mathcal{V}}^{-1}$) with distance and similar oscillation pattern originating from the antiferromagnetic correlations in the spin chain (see the inset in Fig. \ref{fig5}). The good linearity confirms clearly the exponential decay of the oscillation amplitude and provides a further support for our analysis for $n^c<1$. We obtain $\xi_{\mathcal{V}}=3$ for $J_K/t=0.8$ and $\xi_{\mathcal{V}}=1$ for $J_K/t=1.6$. No oscillation is observed for larger $J_K/t$. This indicates that $J_K/t=1.6$ locates at the boundary of weak and intermediate couplings for $n_c=1$. The very rapid suppression of the oscillation at $n^c=1$ suggests that the charge oscillation (for $n^c<1$) plays the role of an enhancer for the hybridization oscillation. The absence of spatial oscillation for larger $J_K/t$ is consistent with our observed $2k_F^{c+f}\,({\rm mod}\,2\pi)$ periodicity for $n^c<1$, but contradicts the mean-field calculations which predicted a wavevector $k=2k_F^c=\pi n^c$ in the hybridization oscillation \cite{Figgins2011}. This difference may provide a possible clue for understanding the discrepancy. In our calculations, the absence of the spatial oscillation for larger $J_K/t$ may be attributed to the effective localization of the redundant conduction electron at the impurity site caused by the large hybridization gap created by the Kondo coupling at its undoped nearest-neighbor sites. The crossover from weak to intermediate couplings reflects the competition between the kinetic energy and the hybridization energy. It would be interesting to check if this charge localization is present in the mean-field calculations.

\section*{Discussion and conclusion}

We note that there has been debate on the exact value of the critical Kondo coupling in the 1D Kondo-Heisenberg model. Different from the DMRG results \cite{Moukouri1996,Eidelstein2011,Xie2015}, some theories argue that the critical Kondo coupling should be at $J_K=0$ \cite{Yamanaka1997}. In this case, there would be no preformed heavy electrons before the formation of the large Fermi surface, but this would seem to be in contradiction with our observed existence of the weak coupling regime. We will not try to solve this difficult issue here due to the limitation of our numerical accuracy in distinguishing the real singularity (rather than a "rapid" change) of the distribution function in the momentum space. In this sense, our observed change at the "critical" Kondo coupling is actually a crossover. Whether or not it will lead to a true singular point in the thermodynamic limit requires more numerical or theoretical scrutiny. For the same reason, we also cannot exclude other possible values of the critical Kondo coupling, although it would then be inconsistent with both numerical tendency \cite{Xie2015} and theoretical argument \cite{Yamanaka1997}. In either case, our numerical calculations yield different predictions from previous mean-field approximation on the oscillation pattern induced by the Kondo holes.

We would like to remark that there may be different understandings of the concept of "heavy electrons". A "standard" definition may depend on the existence of the Fermi liquid state with a large Fermi surface. In this case, heavy electrons are Landau quasiparticles with a large effective mass. However, this definition is limited to a small parameter range in the generic phase diagram. At high temperatures well above the Fermi liquid temperature, or in the antiferromagnetic phase with a small Fermi surface, one also discusses the emergence or existence of heavy electrons, as manifested in a logarithmically divergent specific heat coefficient \cite{Yang2008b} or a large cyclotron mass as measured in the de Haas-van Alphen (dHvA) experiment \cite{Shishido2006}. Especially in 1D, the ground state of the strong coupling limit may not be a Fermi liquid, even if it has a large Fermi surface. Therefore, we adopt in this work a more general "definition" of the concept of heavy electrons to reflect its usual meaning in the literature, namely, a composite state of conduction electrons and localized Heisenberg spins. Unambiguous experimental identification of the composite nature of heavy electrons is difficult due to its small energy scales. Some indirect signatures include the Fano interference in STM or the point-contact spectroscopy as first proposed by one of the authors \cite{Yang2009}, and the hybridization gap in the optical conductivity \cite{Chen2016}. Recent angle-resolved photoemission spectroscopy (ARPES) measurement \cite{Chen2017} and quasiparticle interference experiment \cite{Allan2013} also reveal some convincing evidences. Our work proposes another property of the composite state, namely a hybridization oscillation with a large Fermi wave vector even outside the Fermi liquid regime.

One may wonder if the obtained heavy electron behavior originates from the disturbance of the Kondo holes to the electronic structures of the neighboring sites rather than the intrinsic properties of the pure Kondo lattice. Due to the limitation of the DMRG method, we cannot calculate the spectral properties of the system in this work. However, we have previously calculated the local density of states of a doped periodic Anderson model on a square lattice using the determinant quantum Monte Carlo (DQMC) method \cite{Wei2017} and found that, instead of producing heavy electrons, Kondo holes actually destroy heavy electron state on their neighboring sites. As a result, increasing the number of Kondo holes will lead to a percolation-type transition from the Kondo lattice physics to the single-ion Kondo physics.

To summarize, we use the density matrix renormalization group method to study the perturbation in the local hybridization, the charge density of the conduction electrons and the correlation function of the local spins induced by Kondo holes in the 1D Kondo-Heisenberg model. We find that all three quantities exhibit spatial oscillations whose amplitude decay exponentially with distance away from the Kondo holes. This suggest that the hybridization perturbation due to Kondo holes is different from the usual Friedel oscillation with a power law decay. Our results indicate that the hybridization oscillation is closely related to the oscillations in the conduction electron charge density and the antiferromagnetic spin correlations. At half filling, where the charge density oscillation is suppressed, the antiferromagnetic spin correlations play the major role in determining the hybridization oscillation. Away from half filling, the charge density oscillation plays the dominant role and determines the wavelength and the decay length of the hybridization oscillation. We find three different regimes in the oscillation pattern. In the intermediate and strong coupling regimes, the wavelength is given by the Fermi wavevector of the large Fermi surface, implying the presence of heavy electrons around the large Fermi surface even before it is well formed. Moreover, the derived large decay length indicates that the Kondo lattice physics is highly nonlocal. This excludes any attempt based on local approximations and demands a proper treatment of the nonlocal and collective nature of the lattice hybridization in pursuit of a satisfactory solution to the Kondo lattice problem.


\section*{Acknowledgements}
This work was supported by the National Natural Science Foundation of China (NSFC Grant No. 11522435), the National Key Research and Development Program of China (No. 2017YFA0303103), the State Key Development Program for Basic Research of China (2015CB921303), the Strategic Priority Research Program (B) of the Chinese Academy of Sciences (Grant No. XDB07020200) and the Youth Innovation Promotion Association CAS.

\section*{Author contributions statement}
Y.Y. conceived the idea and supervised the project. N.X., D.H. and Y.Y. performed the research. Y.Y. wrote the paper.

\section*{Additional information}

\textbf{Competing financial interests:} The authors declare no competing financial interests.

\end{document}